\documentclass[vecphys]{svmult}

\usepackage{makeidx}         
\usepackage{graphicx}        
\usepackage{multicol}        
\usepackage[bottom]{footmisc}

\makeindex             

\def\gappeq{\mathrel{\rlap {\raise.5ex\hbox{$>$}} {\lower.5ex\hbox{$\sim$}}}}
\def\lappeq{\mathrel{\rlap{\raise.5ex\hbox{$<$}} {\lower.5ex\hbox{$\sim$}}}}
\begin{document}

\title*{Particle Physics: a Progress Report}

\author{Guido Altarelli}
\institute{
Dipartimento di Fisica `E.~Amaldi', Universit\`a di Roma Tre
and INFN, Sezione di Roma Tre, I-00146 Rome, Italy and \\ CERN, Department of Physics, Theory Division, 
 CH--1211 Geneva 23, Switzerland\\
\texttt{guido.altarelli@cern.ch}}

%
%
\maketitle



\section{Introduction}
\label{sec:1}

I would like to present a concise review of where we stand in particle physics today. First I will discuss QCD, then the electroweak sector and finally the motivations and the avenues for new physics beyond the Standard Model.

\section{QCD}
\label{sec:2}

QCD stands as a main building block of the Standard Model (SM) of particle physics. For many years the relativistic quantum field theory of reference was QED, but at present QCD offers a much more complex and intriguing theoretical laboratory. Indeed, due to asymptotic freedom, QCD can be considered as a better defined theory than QED. The statement that QCD is an unbroken renormalisable gauge theory with six kinds of triplets quarks with given masses completely specifies the form of the Lagrangian in terms of quark and gluon fields. From the compact form of its Lagrangian one might be led to  think that QCD is a "simple" theory. But actually this simple theory has an extremely rich dynamical content, including the property of confinement,  the complexity of the observed hadronic spectrum (with light and heavy quarks), the spontaneous breaking of (approximate) chiral symmetry, a complicated phase transition structure (deconfinement, chiral symmetry restauration, colour superconductivity), a highly non trivial vacuum topology (instantons, $U(1)_A$ symmetry breaking, strong CP violation,....), the property of asymptotic freedom and so on.

How do we get predictions from QCD? There are non perturbative methods: lattice simulations (in great continuous progress), effective lagrangians valid in restricted specified domains [chiral lagrangians, heavy quark effective theories, Soft Collinear Effective Theories (SCET), Non Relativistic QCD....] and also QCD sum rules, potential models (for quarkonium) and so on. But the perturbative approach, based on asymptotic freedom and valid for hard processes, still remains the main quantitative connection to experiment.

Due to confinement no free coloured particles are observed but only colour singlet hadrons. In high energy collisions the produced quarks and gluons materialize as narrow jets of hadrons. Our understanding of the confinement mechanism has much improved thanks to lattice simulations of QCD at finite temperatures and densities \cite{ej}.  The potential between two colour charges clearly shows a linear slope at large distances (linearly rising potential). The slope decreases with increasing temperature until it vanishes at a critical temperature $T_C$. Above $T_C$ the slope remains zero. The phase transitions of colour deconfinement and of chiral restauration appear to happen together on the lattice. Near the critical temperature for both deconfinement and chiral restauration a rapid transition is observed in lattice simulations. In particular the energy density $\epsilon(T)$ is seen to sharply increase. The critical parameters and the nature of the phase transition depend on the number of quark flavours $N_f$ and on their masses. For example, for  $N_f$ = 2 or 2+1 (i.e. 2 light u and d quarks and 1 heavier s quark), $T_C \sim 175~MeV$  and $\epsilon(T_C) \sim 0.5-1.0 ~GeV/fm^3$. For realistic values of the masses $m_s$ and $m_{u,d}$ the phase transition appears to be a second order one, while it becomes first order for very small or very large $m_{u,d,s}$. The hadronic phase and the deconfined phase are separated by a crossover line at small densities and by a critical line at high densities. Determining the exact location of the critical point in T and $\mu_B$ is an important challenge for theory which is also important for the interpretation of heavy ion collision experiments. At high densities the colour superconducting phase is probably also present with bosonic diquarks acting as Cooper pairs. 

A large investment is being done in experiments of heavy ion collisions with the aim of finding some evidence of the quark gluon plasma phase. Many exciting results have been found at the CERN SPS in the past  years and more recently at RHIC. At the CERN SPS some experimental hints of variation with the energy density were found in the form, for example, of $J/ \Psi$ production suppression or of strangeness enhancement when going from p-A to Pb-Pb collisions. Indeed a posteriori the CERN SPS appears well positioned in energy to probe the transition region, in that a marked variation of different observables was observed. The most impressive effect detected at RHIC, interpreted as due to the formation of a hot and dense bubble of matter, is the observation of a strong suppression of back-to-back correlations in jets from central collisions in Au-Au, showing that the jet that crosses the bulk of the dense region is absorbed. The produced hot matter shows a high degree of collectivity, as shown by the observation of elliptic flow (produced hadrons show an elliptic distribution while it would be spherical for a gas) and resembles a perfect liquid with small or no viscosity. However, for quark gluon plasma, it is fair to say that the significance of each single piece of evidence can be questioned and one is still far from an experimental confirmation of a phase transition. The experimental programme on heavy ion collisions will continue at RHIC and then at the LHC where ALICE, a dedicated heavy ion collision experiment, is in preparation.

As we have seen, a main approach to non perturbative problems in QCD is by simulations of the theory on the lattice, a technique started by K. Wilson in 1974 which has shown continuous progress over the last decades. One recent big step, made possible by the availability of more powerful dedicated computers, is the evolution from quenched (i.e. with no dynamical fermions) to unquenched calculations. In doing so an evident improvement in the agreement of predictions with the data is obtained. For example \cite{kro}, modern unquenched simulations reproduce the hadron spectrum quite well. Calculations with dynamical fermions (which take into account the effects of virtual quark loops) imply the evaluation of the quark determinant which is a difficult task. How difficult depends on the particular calculation method. There are several approaches (Wilson, twisted mass,  Kogut-Susskind staggered, Ginsparg-Wilson fermions), each with its own advantages and disadvantages (including the time it takes to run the simulation on a computer). Another area of progress is the implementation of chiral extrapolations: lattice simulation is limited to large enough masses of light quarks. To extrapolate the results down to the physical pion mass one can take advantage of the chiral effective theory in order to control the chiral logs: $\log(m_q/4\pi f_\pi)$. For lattice QCD one is now in an epoch of pre-dictivity as opposed to the  post-dictivity of the past. And in fact the range of precise lattice results currently includes many domains:  the QCD coupling constant (the value $\alpha_s(m_Z)=0.1170(12)$ has been recently quoted \cite{alf}: the central value is in agreement with other determinations but one would not trust the stated error as the total uncertainty), the quark masses, the form factors for K and D decay, the B parameter for kaons, the decay constants $f_K$, $f_D$, $f_{Ds}$, the $B_c$ mass, the nucleon axial charge $g_A$ (the lattice result \cite{neg} is close to the experimental value $g_A \sim 1.25$ and well separated from the $SU(6)$ value $g_A = 5/3$) and many more.

Recently some surprising developments in hadron spectroscopy have attracted the general attention. Ordinary hadrons are baryons, $B\sim qqq$ and mesons $M\sim q \bar q$. For a long time the search for exotic states was concentrated on glueballs, gg bound states, predicted  at $M\gappeq 1.5 ~GeV$ by the lattice. As well known, experimentally glueballs were never clearly identified, probably because they are largely mixed with states made up of quark-antiquark pairs. Hybrid states ($q \bar q g$ or $qqqg$) have also escaped detection. Recently a number of unexpected results have revamped the interest for hadron spectroscopy. Several experiments have reported new narrow states, with widths below a few MeV(!!): 
$\Theta^+(1540)$ with the quantum numbers of $nK^+$ or $pK^0_S$ or, in terms of quarks, of $uudd \bar s$; $D_{sJ}^+(2317) \sim D_s\pi$, $D_{sJ}^+(2460) \sim D^*_s\pi$,.... and $X^0(3872) \sim \pi \pi J/\Psi$. The interpretations proposed are in terms of pentaquarks ($[ud][ud]\bar s$ for $\Theta^+$ for example), tetraquarks ($[qq][\bar q \bar q]$) vs meson-meson molecules for low lying scalar mesons or for 
$X^0$ and also in terms of chiral solitons. Tetraquarks and pentaquarks are based on diquarks: $[qq]$
of spin $0$, antisymmetric in colour, $\bar 3$ of $SU(3)_{colour}$, and antisymmetric in flavour, $\bar 3$ of $SU(3)_{flavour}$. Tetraquarks were originally proposed for scalar mesons by Jaffe \cite{jaf}. It is well known that there are two clusters of scalar mesons: one possible nonet at high mass, around $1.5~GeV$, and a low lying nonet below $1~GeV$. The light nonet presents an inversion in the spectrum: the mesons that would contain  s-quarks in the conventional $q\bar q$ picture and would hence be heavier are actually lighter. In the tetraquark interpretation this becomes clear because the s quark with index "3" of the conventional picture is now replaced be the diquark $[ud]$. However, one can still formulate doubts about the existence of so many scalar states \cite{pen}. The tetraquark interpretation for the doubly charmed $X^0(3872)$ has been proposed recently by Maiani et al \cite{mai} as opposed to that in terms of a $D-D*$ molecule by Braaten and Kusunoki \cite{bra}. Both models appear to face difficulties with the data. For putative pentaquark states like the $\Theta^+$ doubts on their existence have much increased recently. Not only there are mass inconsistencies among different experiments,  evident tension between a small width and large production rates and the need of an exotic production mechanism to explain the lack of evidence at larger energies. But the most disturbing fact is the absence of the signal in some specific experiments where it is difficult to imagine a reason for not seeing it \cite{kle}.

We  now discuss  perturbative QCD \cite{gal}. In the QCD Lagrangian quark masses are the only parameters with dimensions. Naively (or classically) one would expect massless QCD to be scale invariant so that dimensionless observables would not depend on the absolute energy scale but only on ratios of energy variables. While massless QCD in the quantum version, after regularisation and renormalisation, is finally not scale invariant, the theory is asymptotically free and all the departures from scaling are asymptotically small, logarithmic and computable in terms of the running coupling $\alpha_s(Q^2)$. Mass corrections, present in the realistic case together with hadronisation effects, are suppressed by powers. The QCD beta function that fixes the running coupling is known in QCD up to 4 loops in the $MS$ or $\bar{MS}$ definitions and the expansion is well behaved. The 4-loop calculation  by van Ritbergen, Vermaseren and Larin \cite{van} involving about 50.000 4-loop diagrams is a great piece of work. The running coupling is a function of $Q^2/\Lambda^2_{QCD}$, where $\Lambda_{QCD}$ is the scale that breaks scale invariance in massless QCD. Its value in $\bar{MS}$, for 5 flavours of quarks, from the PDG'06 is $\Lambda_{QCD}\sim 222(25)~MeV$. This fundamental constant of nature, which determines the masses of hadrons, is a subtle effect arising from defining the theory at the quantum level. There is no hierarchy problem in QCD, in that the logarithmic evolution of the running makes the smallness of $\Lambda_{QCD}$ with respect to the Planck mass $M_{Pl}$ natural: $\Lambda_{QCD}\sim M_{Pl} \exp{[-1/2b\alpha_s(M_{Pl}^2)]}$. 

The measurements of $\alpha_s(Q^2)$  are among the main quantitative tests of the theory. The most  precise and  reliable determinations are from $e^+e^-$ colliders (mainly at LEP: inclusive hadronic Z decay, inclusive
hadronic $\tau$ decay, event shapes and jet rates) and from scaling violations in Deep Inelastic Scattering (DIS).  Z decay widths are very clean: the perturbative expansion is known to 3-loops, power  corrections are controlled by the light-cone operator expansion and are very suppressed due to $m_Z$ being very large. For measuring $\alpha_s(Q^2)$ \cite{ste} the basic quantity is $\Gamma_h$ the Z hadronic partial width. It enters in $R_l$, $\sigma_h$, $\sigma_l$ and $\Gamma_Z$ (the width ratio of hadrons to leptons, the hadron cross section at the peak, the charged lepton cross section at the peak and the total width, respectively) which are separately measured with largely independent systematics. From combining all these measurements one obtains $\alpha_s(m^2_Z)= 0.1186(27)$ \cite{lew}. The error is predominantly theoretical and is dominated by our ignorance on $m_H$ and from higher orders in the QCD expansion (the possible impact of new physics is very limited, given the results of precision tests of the SM at LEP).  The measurement of $\alpha_s(m_Z)$ from $\tau$ decay is based on $R_\tau$, the ratio of the hadronic to leptonic widths. $R_\tau$ has a number of advantages that, at least in part, tend to compensate for the smallness of $m_\tau$. First, $R_\tau$ is maximally inclusive, more than $R_{e^+e^-}(s)$, because one also integrates over all values of the invariant hadronic squared mass. Analyticity is used to transform the integral into one on the circle at $|s|=m_\tau^2$.  Also, a factor $(1-\frac{s}{m_\tau^2})^2$ that appears in the integral kills the sensitivity of the region $\rm{Re}(s)=m_\tau^2$ where the physical cut and the associated thresholds are located. Still the quoted result (PDG'06) looks a bit too precise: $\alpha_s(m_Z^2)=0.120(3)$.
This precision is obtained by taking for granted that corrections suppressed by $1/m_\tau^2$ are negligible. This is
because, in the massless theory, no dim-2 Lorentz and gauge invariant operators exist that can appear in the light cone expansion. In
the massive theory, the coefficient of $1/m_\tau^2$ does not vanish but is proportional to light quark mass-squared $m^2$. This is still negligible if $m$ is taken as a Lagrangian mass of a few $MeV$. But would not at all be negligible, actually would much increase the theoretical error, if it is taken as a constituent mass of order $m\sim\Lambda_{QCD}$. Most people believe the optimistic version. I am not convinced that the gap is not filled up by ambiguities of $0(\Lambda_{QCD}^2/m_\tau^2)$ e.g. from ultraviolet renormalons. In any case, one can discuss the error, but it is true and remarkable, that the central value from $\tau$ decay, obtained at very small $Q^2$, when evolved at $Q^2=m_Z^2$, is in perfect agreement with all other precise determinations of $\alpha_s(m_Z^2)$ at more typical LEP values of $Q^2$. The measurements of $\alpha_s$ from event shapes and jet rates are affected by non perturbative hadronic corrections which are difficult to precisely assess. The combined result gives $\alpha_s(m_Z^2)=0.120(5)$ (PDG'06).  By measuring event shapes at different energies in the LEP1 and LEP2 ranges one also directly sees the running of $\alpha_s$.

In DIS QCD predicts the $Q^2$ dependence of a generic structure function $F(x,Q^2)$ at each fixed x, not the x shape. But the $Q^2$ dependence is related to the x shape by the QCD evolution equations. For each x-bin the data allow to extract the slope of an approximately straight line, the log slope:  $dlogF(x,Q^2)/dlogQ^2$.  For most x values the $Q^2$ span and the precision of the data are not much sensitive to the curvature. A single value of $\Lambda_{QCD}$ must be fitted to reproduce the collection of the log slopes. The QCD theory of scaling violations, based on the renormalization group and the light-cone operator expansion, is crystal clear. Recently ('04) the formidable task of computing the splitting functions at NNLO accuracy has been completed by Moch, Vermaseren and Vogt, a really monumental, fully analytic calculation \cite{moc}. For the determination of $\alpha_s$ the scaling violations of non-singlet structure functions would be ideal, because of the minimal impact of the choice of input parton densities. Unfortunately the data on non-singlet structure functions are not very accurate. For example, NNLO determinations of $\alpha_s$ from the CCFR data on $F_{3\nu N}$ with different techniques have led to the central values $\alpha_s(m_Z^2)=0.1153$ \cite{san}), $\alpha_s(m_Z^2)=0.1174$ \cite{max}, $\alpha_s(m_Z^2)=0.1190$ \cite{kat}, with average and common estimated error of $\alpha_s(m_Z^2)=0.117(6)$ which I will use later. When one measures $\alpha_s$ from scaling violations on $F_2$ from e or $\mu$ beams, the data are abundant, the errors small but there is an increased dependence on input parton densities and especially a strong correlation between the result on $\alpha_s$ and the input on the gluon density. There are several most complete and accurate derivations of $\alpha_s$ from scaling violations in $F_2$ with different, sophisticated methods (Mellin moments, Bernstein moments, truncated moments....).  We quote here the result at NNLO accuracy from MRST'04 (see PDG'06): $\alpha_s(m_Z^2)=0.1167(40)$.

More measurements of $\alpha_s$ could be listed: I just reproduced those which I think are most significant and reliable. There is a remarkable agreement among the different determinations. If I directly average the five values listed above from inclusive Z decay, from $R_\tau$, from event shapes and jet rates in $e^+e^-$, from $F_3$ and from $F_2$ in DIS I obtain $\alpha_s(m_Z^2)=0.1187(16)$ in good agreement with the PDG'06 average $\alpha_s(m_Z^2)=0.1176(20)$.

The importance of DIS for QCD goes well beyond the measurement of $\alpha_s$. In the past it played a crucial role in establishing the reality of quarks and gluons as partons and in promoting  QCD as the theory of strong interactions. Nowadays it still generates challenges to QCD as, for example, in the domain of structure functions at small x or of polarized structure functions or of generalized parton densities and so on. 

The problem of constructing a convergent procedure to include the BFKL corrections at small x in the singlet splitting functions, in agreement with the small-x behaviour observed at HERA, has been a long standing puzzle which has now been essentially solved. The naive BFKL rise of splitting functions is tamed by resummation of collinear singularities and by running coupling effects. The resummed expansion is well behaved and the result is close to the perturbative NLO splitting function in the region of HERA data at small x \cite{abf},\cite{ccss}. 

In polarized DIS one main question is how the proton helicity is distributed among quarks, gluons and orbital angular momentum: $1/2\Delta \Sigma + \Delta g + L_z= 1/2$ \cite{spi}.
The quark moment $\Delta \Sigma$ was found to be small: typically, at $Q^2\sim 1 GeV^2$, $\Delta \Sigma_{exp} \sim 0.2$ (the "spin crisis"). Either $\Delta g + L_z$ is large or there are contributions to $\Delta \Sigma$ at very small x outside of the measured region. $\Delta g$ evolves like $\Delta g \sim log Q^2$, so that eventually should become large (while $\Delta \Sigma$ and $\Delta g + L_z$ are $Q^2$ independent in LO). It will take long before this log growth of $\Delta g$ will be confirmed by experiment! $\Delta g$ can be measured indirectly by scaling violations and directly from asymmetries, e.g. in $c \bar c$ production. Existing direct measurements by Hermes, Compass, and at RHIC are still very crude and show no hint of a large $\Delta g$. The perspectives of better measurements are good at Compass and RHIC in the near future.

Another important role of DIS is to provide information on parton density functions (PDF) which are instrumental for computing cross-sections of hard processes at hadron colliders via the factorisation formula. The predictions for cross sections and distributions at $pp$ or $p\bar p$ colliders for large $p_T$ jets or photons, for heavy quark production, for Drell-Yan, W and Z production are all in very good agreement with experiment. There was an apparent problem for b quark production at the Tevatron, but the problem appears now to be solved by a combination of refinements (log resummation, B hadrons instead of b quarks, better fragmentation functions....)\cite{cac}. The QCD predictions are so solid that W and Z production are actually considered as possible luminosity monitors for the LHC. 

A great effort is being devoted to the preparation to the LHC. Calculations for specific processes are being completed. A very important example is Higgs production via $g ~+~ g \rightarrow H$. The amplitude is dominated by the top quark loop. Higher order corrections can be computed either in the effective lagrangian approach, where the heavy top is integrated away and the loop is shrunk down to a point [the coefficient of the effective vertex is known to $\alpha_s^4$ accuracy \cite{che}], or in the full theory. At the NLO \cite{sti} the two approaches agree very well for the rate as a function of $m_H$. Rapidity and $p_T$ distributions have also been evaluated at NLO \cite{sti}. The $[log(p_T/m_H)]^n$ have been resummed in analogy with what was done long ago for W and Z production. Recently the NNLO analytic calculation for the rate has been completed in the effective lagrangian formalism \cite{sti}, \cite{ana}. 

The activity on event simulation also received a big boost from the LHC preparation. General algorithms for performing NLO calculations numerically  (requiring techniques for the cancellation of singularities between real and virtual diagrams), for example the dipole formalism by Catani, Seymour et al. \cite{cat}, have been developed. The matching of matrix element calculation of rates together with the modeling of parton showers has been realised in packages, as for example in the MC@NLO based on HERWIG. The matrix element calculation, improved by resummation of large logs, provides the hard skeleton (with large $p_T$ branchings) while the parton shower is constructed by a sequence of factorized collinear emissions fixed by the QCD splitting functions. In addition, at low scales a model of hadronisation completes the simulation. The importance of all the components, matrix element, parton shower and hadronisation can be appreciated in simulations of hard events compared with the Tevatron data. 

Before closing I would like to mention some very interesting developments at the interface between string theory and QCD, twistor calculus. A precursor work was the Parke-Taylor result in 1986 \cite{par} on the amplitudes for n  gluons (all taken as incoming) with given ± helicities. Inspired by dual models, they derived a compact formula for the maximum non vanishing helicity violating amplitude (with n-2  plus and 2 minus helicities) in terms of spinor products. Using the relation between strings and gauge theories in twistor space Witten \cite{wit} developed in '03 a formalism in terms of effective vertices and propagators that allows to compute all helicity amplitudes. The method, much faster than Feynman diagrams, leads to very compact results. Since then rapid progress followed \cite{sti}: for tree level processes powerful recurrence relations were established (Britto, Cachazo, Feng; Witten), the method was extended  to include massless fermions (Georgiu, Khoze) and also external EW vector bosons (Bern et al)  and Higgs particles (Dixon, Glover, Khoze, Badger et al). The level already attained is already important for multijet events at the LHC. And the study of loop diagrams has been started. In summary, this road looks very promising. 

A different string connection is the attempt at obtaining results on QCD from the AdS correspondence, pioneered by Maldacena \cite{mal} . The starting point is the holographic correspondence between $D=10$ string theory and the $N=4$ SUSY Yang-Mills in four dimensions at large $N_c$. From there to get to real life QCD the way looks impervious, but a number of results for actual processes have  been advocated and the perspective is exciting \cite{sho}.

In conclusion, I think that the domain of QCD appears as one of great maturity but also of robust vitality with many rich branches and plenty of new blossoms. The physics content of QCD is very large and our knowledge, especially in the non perturbative domain, is still very limited but progress both from experiment (LEP, HERA, Tevatron, RHIC, LHC......) and from theory is continuing at a healthy rate. And all the QCD predictions that we were able to formulate and to test are in very good agreement with experiment.

\section{The Physics of Flavour}
\label{sec:3}

In the last decade great progress in different areas of flavour physics has been achieved. In the quark sector, the amazing results of a generation of frontier experiments, obtained at B factories and at accelerators, have become available. QCD has been playing a crucial role in the interpretation of experiments by a combination of effective theory methods (heavy quark effective theory, NRQCD, SCET), lattice simulations and perturbative calculations. The hope of these experiments was to detect departures from the CKM picture of mixing and CP violation as  signals of new physics. Finally, B mixing and CP violation agree very well with the SM predictions based on the CKM matrix \cite{fle}. The recent measurement of $\Delta m_s$ by CDF and D0, in fair agreement with the SM expectation, has closed another door for new physics. It is only in channels that are forbidden at tree level and occur through penguin loops (as is the case for $B \rightarrow  \pi K$ modes) that some deviation could be hidden. The amazing performance of the SM in flavour changing transitions and for CP violation in K and B decays poses a strong constraint on all proposed models of new physics. 

In the leptonic sector the study of neutrino oscillations has led to the discovery that at least two neutrinos are not massless and to the determination of the mixing matrix \cite{alfe}. Neutrinos are not all massless but their masses are very small. Probably masses are small because $\nu$Õs are Majorana particles, and, by the see-saw mechanism, their masses are inversely proportional to the large scale $M$ where lepton number ($L$) violation occurs (as expected in GUT's). Indeed the value of $M\sim m_{\nu R}$ from experiment is compatible with being close to $M_{GUT} \sim 10^{14}-10^{15}GeV$, so that neutrino masses fit well in the GUT picture and actually support it.  It was realized that decays of heavy $\nu_R$ with CP and L violation can produce a B-L asymmetry. The range of neutrino masses indicated by neutrino phenomenology turns out to be perfectly compatible with the idea of baryogenesis via leptogenesis \cite{buch}. This elegant model for baryogenesis has by now replaced the idea of baryogenesis near the weak scale, which has been strongly disfavoured by LEP. It is remarkable that we now know the neutrino mixing matrix with good accuracy. Two mixing angles are large and one is small. The atmospheric angle $\theta_{23}$ is large, actually compatible with maximal but not necessarily so: at $3\sigma$: $0.31 \leq \sin^2{\theta_{23}}\leq 0.72$ with central value around  $0.5$. The solar angle $\theta_{12}$ is large, $\sin^2{\theta_{12}}\sim 0.3$, but certainly not maximal (by more than 5$\sigma$). The third angle $\theta_{13}$, strongly limited mainly by the CHOOZ experiment, has at present a $3\sigma$ upper limit given by about $\sin^2{\theta_{13}}\leq 0.08$. While these discoveries are truly remarkable, it is somewhat depressing that the detailed knowledge of both the quark and the neutrino mixings has not led so far to a compelling solution of the dynamics of fermion masses and mixings: our models can reproduce, actually along different ways, the observed values, but we do not really understand their mysterious pattern.

\section{Precision Tests of the Standard Electroweak Theory}
\label{sec:4}

The results of the electroweak precision tests as well as of the searches
for the Higgs boson and for new particles performed at LEP and SLC are now available in  final form.  Taken together
with the measurements of $m_t$, $m_W$ and the searches for new physics at the Tevatron, and with some other data from low
energy experiments, they form a very stringent set of precise constraints \cite{lew} to compare with the Standard Model (SM) or with
any of its conceivable extensions. When confronted with these results, on the whole the SM performs rather
well, so that it is fair to say that no clear indication for new physics emerges from the data \cite{AG}. The main lesson of precision tests of the standard electroweak theory
can be summarised as follows. The couplings of quark and leptons to
the weak gauge bosons W$^{\pm}$ and Z are indeed precisely those
prescribed by the gauge symmetry. The accuracy of a few per-mille for
these tests implies that, not only the tree level, but also the
structure of quantum corrections has been verified. To a lesser
accuracy the triple gauge vertices $\gamma WW$ and Z$WW$ have also
been found in agreement with the specific prediction of the
$SU(2)\bigotimes U(1)$ gauge theory. This means that it has been
verified that the gauge symmetry is unbroken in the vertices of the
theory: all currents and charges are indeed symmetric. Yet there is obvious
evidence that the symmetry is otherwise badly broken in the
masses. This is a clear signal of spontaneous
symmetry breaking. The practical implementation of spontaneous
symmetry breaking in a gauge theory is via the Higgs mechanism. The
Higgs sector of the SM is still very much untested. What has been
tested is the relation $M_W^2=M_Z^2\cos^2{\theta_W}$, modified by small, computable
radiative corrections. This relation means that the effective Higgs
(be it fundamental or composite) is indeed a weak isospin doublet.
The Higgs particle has not been found but in the SM its mass can well
be larger than the present direct lower limit $m_H \gappeq 114$~GeV
obtained from direct searches at LEP-2.  The radiative corrections
computed in the SM when compared to the data on precision electroweak
tests lead to a clear indication for a light Higgs, not too far from
the present lower bound. The exact upper limit for $m_H$ in the SM depends on the value of the top quark mass $m_t$ (the one-loop radiative corrections are quadratic in $m_t$ and logarithmic in $m_H$). The measured value of $m_t$ went down recently (as well as the associated error) according to the results of Run II at the Tevatron. The CDF and D0 combined value is at present \cite{gle} $m_t~= 171.4~\pm~2.1~GeV$ (it went slightly down with respect to the value from Run I). As a consequence the present limit on $m_H$ is more stringent \cite{woo}: $m_H < 199~GeV$ (at $95\%$ c.l., after including the information from the 114 GeV direct bound).

\section{Outlook on Avenues beyond the Standard Model}
\label{sec:5}

No signal of new physics has been
found neither in electroweak precision tests nor in flavour physics. Given the success of the SM why are we not satisfied with that theory? Why not just find the Higgs particle,
for completeness, and declare that particle physics is closed? The reason is that there are
both conceptual problems and phenomenological indications for physics beyond the SM. On the conceptual side the most
obvious problems are that quantum gravity is not included in the SM and the related hierarchy problem. Among the main
phenomenological hints for new physics we can list coupling unification, dark matter, neutrino masses, 
baryogenesis and the cosmological vacuum energy.

The computed evolution with energy
of the effective SM gauge couplings clearly points towards the unification of the electro-weak and strong forces (Grand
Unified Theories: GUT's) at scales of energy
$M_{GUT}\sim  10^{15}-10^{16}~ GeV$ which are close to the scale of quantum gravity, $M_{Pl}\sim 10^{19}~ GeV$.  One is led to
imagine  a unified theory of all interactions also including gravity (at present superstrings provide the best attempt at such
a theory). Thus GUT's and the realm of quantum gravity set a very distant energy horizon that modern particle theory cannot
ignore. Can the SM without new physics be valid up to such large energies? One can imagine that some obvious problems could be postponed to the more fundamental theory at the Planck mass. For example, the explanation of the three generations of fermions and the understanding of fermion masses and mixing angles can be postponed. But other problems must find their solution in the low energy theory. In particular, the structure of the
SM could not naturally explain the relative smallness of the weak scale of mass, set by the Higgs mechanism at $\mu\sim
1/\sqrt{G_F}\sim  250~ GeV$  with $G_F$ being the Fermi coupling constant. This so-called hierarchy problem is due to the instability of the SM with respect to quantum corrections. This is related to
the
presence of fundamental scalar fields in the theory with quadratic mass divergences and no protective extra symmetry at
$\mu=0$. For fermion masses, first, the divergences are logarithmic and, second, they are forbidden by the $SU(2)\bigotimes
U(1)$ gauge symmetry plus the fact that at
$m=0$ an additional symmetry, i.e. chiral  symmetry, is restored. Here, when talking of divergences, we are not
worried of actual infinities. The theory is renormalisable and finite once the dependence on the cut off $\Lambda$ is
absorbed in a redefinition of masses and couplings. Rather the hierarchy problem is one of naturalness. We can look at the
cut off as a parameterization of our ignorance on the new physics that will modify the theory at large energy
scales. Then it is relevant to look at the dependence of physical quantities on the cut off and to demand that no
unexplained enormously accurate cancellations arise. 

The hierarchy problem can be put in very practical terms: loop corrections to the higgs mass squared are
quadratic in $\Lambda$. The most pressing problem is from the top loop.
 With $m_h^2=m^2_{bare}+\delta m_h^2$ the top loop gives 
 \begin{eqnarray}
\delta m_{h|top}^2\sim -\frac{3G_F}{2\sqrt{2} \pi^2} m_t^2 \Lambda^2\sim -(0.2\Lambda)^2 \label{top}
\end{eqnarray}

If we demand that the correction does not exceed the light Higgs mass indicated by the precision tests, $\Lambda$ must be
close, $\Lambda\sim o(1~TeV)$. Similar constraints arise from the quadratic $\Lambda$ dependence of loops with gauge bosons and
scalars, which, however, lead to less pressing bounds. So the hierarchy problem demands new physics to be very close (in
particular the mechanism that quenches the top loop). Actually, this new physics must be rather special, because it must be
very close, yet its effects are not clearly visible (the "LEP Paradox" \cite{BS}). Examples of proposed classes of solutions
for the hierarchy problem are:

¥ $\bf{Supersymmetry.}$ In the limit of exact boson-fermion symmetry the quadratic divergences of bosons cancel so that
only log divergences remain. However, exact SUSY is clearly unrealistic. For approximate SUSY (with soft breaking terms),
which is the basis for all practical models, $\Lambda$ is replaced by the splitting of SUSY multiplets, $\Lambda\sim
m_{SUSY}-m_{ord}$. In particular, the top loop is quenched by partial cancellation with s-top exchange, so the s-top cannot be too heavy.

¥ $\bf{Technicolor.}$ The Higgs system is a condensate of new fermions. There are no fundamental scalar Higgs sector, hence no
quadratic devergences associated to the $\mu^2$ mass in the scalar potential. This mechanism needs a very strong binding force,
$\Lambda_{TC}\sim 10^3~\Lambda_{QCD}$. It is  difficult to arrange that such nearby strong force is not showing up in
precision tests. Hence this class of models has been disfavoured by LEP, although some special class of models have been devised aposteriori, like walking TC, top-color assisted TC etc (for recent reviews, see, for example, \cite{L-C}). 

$\bf{Large~extra~dimensions.}$ The idea is that $M_{PL}$ appears very large, or equivalently that gravity appears very weak,
because we are fooled by hidden extra dimensions so that the real gravity scale is reduced down to a lower scale, even possibly down to
$o(1~TeV)$. This possibility is very exciting in itself and it is really remarkable that it is compatible with experiment.

¥ $\bf{"Little~Higgs" models.}$ In these models extra symmetries allow $m_h\not= 0$ only at two-loop level, so that $\Lambda$
can be as large as
$o(10~TeV)$ with the Higgs within present bounds (the top loop is quenched by exchange of heavy vectorlike new  quarks with charge 2/3).

We now briefly comment in turn on these possibilities.

SUSY models are the most developed and most widely accepted. Many theorists consider SUSY as established at the Planck
scale $M_{Pl}$. So why not to use it also at low energy to fix the hierarchy problem, if at all possible? It is interesting
that viable models exist. The necessary SUSY breaking can be introduced through soft
terms that do not spoil the good convergence properties of the theory. Precisely those terms arise from
supergravity when it is spontaneoulsly broken in a hidden sector. This is the case of the MSSM \cite{Martin}. Of course, minimality is only a simplicity assumption that could possibly be relaxed. For example, adding an additional Higgs singlet S considerably helps in addressing naturalness constraints \cite{nmssm}, \cite{barbie}. Minimal versions or, even more, very constrained versions like the CMSSM (where simple conditions at the GUT scale are in addition assumed) are economic in terms of new parameters but could be to some extent misleading. Still, the MSSM 
is a completely specified,
consistent and computable theory which is compatible with all precision electroweak tests. In this
most traditional approach SUSY is broken in a hidden sector and the scale of SUSY breaking is very
large of order
$\Lambda\sim\sqrt{G^{-1/2}_F M_{Pl}}$. But since the hidden sector only communicates with the visible sector
through gravitational interactions the splitting of the SUSY multiplets is much smaller, in the TeV
energy domain, and the Goldstino is practically decoupled. 
But alternative mechanisms of SUSY breaking are also being considered. In one alternative scenario \cite{gau} the (not so
much) hidden sector is connected to the visible one by ordinary gauge interactions. As these are much
stronger than the gravitational interactions, $\Lambda$ can be much smaller, as low as 10-100
TeV. It follows that the Goldstino is very light in these models (with mass of order or below 1 eV
typically) and is the lightest, stable SUSY particle, but its couplings are observably large. The radiative
decay of the lightest neutralino into the Goldstino leads to detectable photons. The signature of photons comes
out naturally in this SUSY breaking pattern: with respect to the MSSM, in the gauge mediated model there are typically
more photons and less missing energy. The main appeal of gauge mediated models is a better protection against
flavour changing neutral currents but naturality problems tend to increase. As another possibility it has been
pointed out that there are pure gravity contributions to soft masses that arise from gravity theory anomalies \cite{ano}. In the assumption that these terms are dominant the associated spectrum and phenomenology have been
studied. In this case gaugino masses are proportional to gauge coupling beta functions, so that the gluino is much heavier
than the electroweak gauginos, and the wino is most often the lightest SUSY particle. 

What is really unique to SUSY with respect to all other extensions of the SM listed above is that the MSSM or
other non minimal SUSY models are well defined and computable up to $M_{Pl}$ and, moreover, are not only compatible but actually 
quantitatively supported by coupling unification and GUT's. At present the most direct
phenomenological evidence in favour of supersymmetry is obtained from the unification of couplings in GUTs.
Precise LEP data on $\alpha_s(m_Z)$ and $\sin^2{\theta_W}$ show that
standard one-scale GUTs fail in predicting $\sin^2{\theta_W}$ given
$\alpha_s(m_Z)$ (and $\alpha(m_Z)$) while SUSY GUTs are in agreement with the present, very precise,
experimental results. If one starts from the known values of
$\sin^2{\theta_W}$ and $\alpha(m_Z)$, one finds \cite{LP} for $\alpha_s(m_Z)$ the results:
$\alpha_s(m_Z) = 0.073\pm 0.002$ for Standard GUTs and $\alpha_s(m_Z) = 0.129\pm0.010$ for SUSY~ GUTs
to be compared with the world average experimental value $\alpha_s(m_Z) =0.118\pm0.002$. Another great asset of SUSY GUT's
is that proton decay is much slowed down with respect to the non SUSY case. First, the unification mass $M_{GUT}\sim~\rm{few}~
10^{16}~GeV$, in typical SUSY GUT's, is about 20-30 times larger than for ordinary GUT's. This makes p decay via gauge
boson exchange negligible and the main decay amplitude arises from dim-5 operators with higgsino exchange, leading to a
rate close but still compatible with existing bounds (see, for example,\cite{AFM}). It is also important that SUSY provides an excellent dark matter candidate, the neutralino.  We finally recall that the range of neutrino masses as indicated by oscillation experiments, when interpreted in the see-saw mechanism, point to $M_{GUT}$ and give additional support to GUTs \cite{alfe}.

In spite of all these virtues it is true that the lack of SUSY signals at LEP and the lower limit on $m_H$ pose problems
for the MSSM. The lightest Higgs particle is predicted in the MSSM to be below $m_h\lappeq~135~GeV$. The limit on the SM
Higgs
$m_H\gappeq~114~GeV$ considerably restricts the available parameter space of the MSSM requiring relatively large $\tan\beta$
($\tan\beta\gappeq~2-3$: at tree level $m^2_h=m^2_Z\cos^2{2\beta}$) and rather heavy s-top (the loop corrections
increase with $\log{\tilde{m_t^2}}$). But we have seen that a heavy s-top is unnatural, because it enters quadratically in the radiative corrections to $\delta m_{h|top}^2$. Stringent naturality constraints also follow from imposing that the electroweak
symmetry breaking occurs at the right place: in SUSY models the breaking is induced by the running of the $H_u$ mass
starting from a common scalar mass $m_0$ at $M_{GUT}$. The squared Z mass $m_Z^2$ can be expressed as a linear
combination of the SUSY parameters $m_0^2$, $m_{1/2}^2$, $A^2_t$, $\mu^2$,... with known coefficients. Barring
cancellations that need fine tuning, the SUSY parameters, hence the SUSY s-partners cannot be too heavy. The LEP limits,
in particular the chargino lower bound $m_{\chi+}\gappeq~100~GeV$, are sufficient to eliminate an important region of the
parameter space, depending on the amount of allowed fine tuning. For example, models based on gaugino universality at the
GUT scale are discarded unless a fine tuning by at least a factor of ~20 is not allowed. Without gaugino
universality \cite{kane} the strongest limit remains on the gluino mass: $m_Z^2\sim 0.7~m_{gluino}^2+\dots$ which is still
compatible with the present limit $m_{gluino}\gappeq~200~GeV$.

The non discovery of SUSY at LEP has given further impulse to the quest for new ideas on physics beyond the SM. Large extra
dimensions \cite{Jo} and "little Higgs" \cite{schm} models are  among the most interesting new directions in model building. Large
extra dimension models propose to solve the hierarchy problem by bringing gravity down from $M_{Pl}$ to $m\sim~o(1~TeV)$ where
$m$ is the string scale. Inspired by string theory one assumes that some compactified extra dimensions are sufficiently large
and that the SM fields are confined to a 4-dimensional brane immersed in a d-dimensional bulk while gravity, which feels the
whole geometry, propagates in the bulk. We know that the Planck mass is large because gravity is weak: in fact $G_N\sim
1/M_{Pl}^2$, where
$G_N$ is Newton constant. The idea is that gravity appears so weak because a lot of lines of force escape in extra
dimensions. Assume you have $n=d-4$ extra dimensions with compactification radius $R$. For large distances, $r>>R$, the
ordinary Newton law applies for gravity: in natural units $F\sim G_N/r^2\sim 1/(M_{Pl}^2r^2)$. At short distances,
$r\lappeq R$, the flow of lines of force in extra dimensions modifies Gauss law and $F^{-1}\sim m^2(mr)^{d-4}r^2$. By
matching the two formulas at $r=R$ one obtains $(M_{Pl}/m)^2=(Rm)^{d-4}$. For $m\sim~1~TeV$ and $n=d-4$ one finds that
$n=1$ is excluded ($R\sim 10^{15} cm$), for $n=2~R$  is at the edge of present bounds $R\sim~1~ mm$, while for $n=4,6$,
$R\sim~10^{-9}, 10^{-12}~cm$.  In all these models a generic feature is the occurrence of Kaluza-Klein (KK) modes.
Compactified dimensions with periodic boundary conditions, as for quantization in a box, imply a discrete spectrum with
momentum $p=n/R$ and mass squared $m^2=n^2/R^2$. There are many versions of these models. The SM brane can itself have a
thickness $r$ with $r<\sim~10^{-17}~cm$ or $1/r>\sim~1~TeV$, because we know that quarks and leptons are pointlike down to
these distances, while for gravity there is no experimental counter-evidence down to $R<\sim~0.1~mm$ or
$1/R>\sim~10^{-3}~eV$. In case of a thickness for the SM brane there would be KK recurrences for SM fields, like $W_n$,
$Z_n$ and so on in the $TeV$ region and above. There are models with factorized metric ($ds^2=\eta_{\mu
\nu}dx^{\mu}dx^{\nu}+h_{ij}(y)dy^idy^j$, where y (i,j) denotes the extra dimension coordinates (and indices), or models
with warped metric ($ds^2=e^{-2kR|\phi|} \eta_{\mu \nu}dx^{\mu}dx^{\nu}-R^2\phi^2$ \cite{RS}.
In any case there are the towers of KK recurrences of the graviton. They
are gravitationally coupled but there are a lot of them that sizably couple, so that the net result is a modification
of cross-sections and the presence of missing energy. 

Large extra dimensions provide a very exciting scenario \cite{FeAa}. Already it is remarkable that this possibility is
compatible with experiment. However, there are a number of criticisms that can be brought up. First, the hierarchy problem is
more translated in new terms rather than solved. In fact the basic relation $Rm=(M_{Pl}/m)^{2/n}$ shows that $Rm$, which one
would apriori expect to be $0(1)$, is instead ad hoc related to the large ratio $M_{Pl}/m$.  In this respect  the Randall-Sundrum variety is more appealing because the hierarchy suppression $m_W/M_{Pl}$ could arise from the warping factor $e^{-2kR|\phi|}$, with not too large values of $kR$. The question of whether these values of $kR$ are reasonable has been discussed in ref. \cite{GW}, which offer the best support to the solution of the hierarchy problem in this context. Also it is
not clear how extra dimensions can by themselves solve the LEP paradox (the large top loop corrections should be
controlled by the opening of the new dimensions and the onset of gravity): since
$m_H$ is light
$\Lambda\sim 1/R$ must be relatively close. But precision tests put very strong limits on $\Lambda$. In fact in typical
models of this class there is no mechanism to sufficiently quench the corrections. While no simple, realistic model has
yet emerged as a benchmark, it is attractive to imagine that large extra dimensions could be a part of the truth,
perhaps coupled with some additional symmetry or even SUSY. The Randall-Sundrum warped geometry has become the common framework for many attempts in this direction.

In the  general context of extra dimensions an interesting direction of development is the study of symmetry breaking by orbifolding and/or boundary conditions. These are models where a larger gauge symmetry (with or without SUSY) holds in the bulk. The symmetry is reduced in the 4 dimensional brane, where the physics that we observe is located, as an effect of symmetry breaking induced geometrically by suitable boundary conditions. There are models where SUSY, valid in $n>4$ dimensions is broken by boundary conditions \cite{ant}, in particular the model of ref.\cite{bar}, where the mass of the Higgs is computable and can be estimated with good accuracy. Then there are "Higgsless  models" where it is the SM electroweak gauge symmetry which is broken at the boundaries \cite{Hless}.  Or models where the Higgs is the 5th component of a gauge boson of an extended  symmetry valid in $n>4$ \cite{hoso}. In general all these alternative models for the Higgs mechanism face severe problems and constraints from electroweak precision tests \cite{BPR}. At the GUT scale, symmetry breaking by orbifolding can be applied to obtain a reformulation of SUSY GUT's where many problematic features of ordinary GUT's (e.g. a baroque Higgs sector, the doublet-triplet splitting problem, fast proton decay etc) are improved \cite{Kaw}, \cite{FeAa}.

In "little Higgs" models the symmetry of the SM is extended to a suitable global group G that also contains some
gauge enlargement of $SU(2)\bigotimes U(1)$, for example $G\supset [SU(2)\bigotimes U(1)]^2\supset SU(2)\bigotimes
U(1)$. The Higgs particle is a pseudo-Goldstone boson of G that only takes mass at 2-loop level, because two distinct
symmetries must be simultaneously broken for it to take mass,  which requires the action of two different couplings in
the same diagram. Then in the relation
between
$\delta m_h^2$ and
$\Lambda^2$ there is an additional coupling and an additional loop factor that allow for a bigger separation between the Higgs
mass and the cut-off. Typically, in these models one has one or more Higgs doublets at $m_h\sim~0.2~TeV$, and a cut-off at
$\Lambda\sim~10~TeV$. The top loop quadratic cut-off dependence is partially canceled, in a natural way guaranteed by the
symmetries of the model, by a new coloured, charge-2/3, vectorial quark $\chi$ of mass around $1~TeV$ (a fermion not a scalar
like the s-top of SUSY models). Certainly these models involve a remarkable level of group theoretic virtuosity. However, in
the simplest versions one is faced with problems with precision tests of the SM \cite{prob}. Even with
vectorlike new fermions large corrections to the epsilon parameters arise from exchanges of the new gauge bosons
$W'$ and $Z'$ (due to lack of custodial $SU(2)$ symmetry). In order to comply with these constraints the cut-off must be
pushed towards large energy and the amount of fine tuning needed to keep the Higgs light is still quite large.
Probably these bad features can be fixed by some suitable complication of the model (see for example, \cite{Ch}). But, in my opinion, the real limit of
this approach is that it only offers a postponement of the main problem by a few TeV, paid by a complete loss of
predictivity at higher energies. In particular all connections to GUT's are lost. An interesting model that combines the idea of the Higgs as a Goldstone boson and warped extra dimensions was proposed and studied in refs.\cite{con}.

Finally, we stress the importance of  the dark matter and of the cosmological constant or vacuum energy problem \cite{tu}. In fact, we know by now \cite{WMAP} that  the  Universe is flat and most of it is not made up of known forms of matter: $\Omega_{tot} \sim 1$, $\Omega_{baryonic} \sim 0.044$, $\Omega_{matter} \sim 0.27$, where $\Omega$ is the ratio of the density to the critical density. Most is Dark Matter (DM) and Dark Energy (DE). We also know that most of DM must be cold (non relativistic at freeze-out) and that significant fractions of hot DM are excluded. Neutrinos are hot DM (because they are ultrarelativistic at freeze-out) and indeed are not much cosmo-relevant: $\Omega_{\nu} \lappeq 0.015$. Identification of DM is a task of enormous importance for both particle physics and cosmology. If really neutralinos are the main component of DM they will be discovered at the LHC and this will be a great service of particle physics to cosmology. More in general, the LHC is sensitive to a large variety of WIMP's (Weekly Interacting Massive Particles). WIMP's with masses in the 10 GeV-1TeV range with typical electroweak crosssections contribute to $\Omega$ terms of $o(1)$.
Also, these results
on cosmological parameters have shown that vacuum energy accounts
for about 2/3 of the critical density: $\Omega_{\Lambda}\sim 0.65$, Translated into familiar units this means for the energy
density $\rho_{\Lambda}\sim (2~10^{-3}~eV)^4$ or $(0.1~mm)^{-4}$. It is really interesting (and not at all understood)
that $\rho_{\Lambda}^{1/4}\sim \Lambda_{EW}^2/M_{Pl}$ (close to the range of neutrino masses). It is well known that in
field theory we expect $\rho_{\Lambda}\sim \Lambda_{cutoff}^4$. If the cut off is set at $M_{Pl}$ or even at $0(1~TeV)$
there would an enormous mismatch. In exact SUSY $\rho_{\Lambda}=0$, but SUSY is broken and in presence of breaking 
$\rho_{\Lambda}^{1/4}$ is in general not smaller than the typical SUSY multiplet splitting. Another closely related
problem is "why now?": the time evolution of the matter or radiation density is quite rapid, while the density for a
cosmological constant term would be flat. If so, then how comes that precisely now the two density sources are
comparable? This suggests that the vacuum energy is not a cosmological constant term, buth rather the vacuum expectation
value of some field (quintessence) and that the "why now?" problem is solved by some dynamical coupling of the quintessence field with gauge singlet  fields (perhaps RH neutrinos). 

Clearly the cosmological constant problem poses a big question mark on the relevance of naturalness as a relevant criterion also for the hierarchy problem: how we can trust that we need new physics close to the weak scale out of naturalness if we have no idea on the solution of the cosmological constant huge naturalness problem? The common answer is that the hierarchy problem is formulated within a well defined field theory context while the cosmological constant problem makes only sense within a theory of quantum gravity, that there could be modification of gravity at the sub-eV scale, that the vacuum energy could flow in extra dimensions or in different Universes and so on. At the other extreme is the possibility that naturalness is misleading. Weinberg \cite{We} has pointed out that the observed order of magnitude of $\Lambda$ can be successfully reproduced as the one necessary to allow galaxy formation in the Universe. In a scenario where new Universes are continuously produced we might be living in a very special one (largely fine-tuned) but the only one to allow the development of an observer (anthropic principle). One might then argue that the same could in principle be true also for the Higgs sector. Recently it was suggested \cite{AHD} to abandon the no-fine-tuning assumption for the electro-weak theory, but require correct coupling unification, presence of dark matter with weak couplings and a single scale of evolution from the EW to the GUT scale. A "split SUSY" model arises as a solution with a fine-tuned light Higgs and all SUSY particles heavy except for gauginos, higgsinos and neutralinos, protected by chiral symmetry. But, then, we could also have a two-scale non-SUSY GUT with axions as dark matter. In conclusion, it is clear that naturalness can be a good heuristic principle but you cannot prove its necessity. The anthropic approach to the hierarchy problem is discussed in ref.s \cite{anto}.

\section{Summary and Conclusion}
\label{sec:6}

Supersymmetry remains the standard way beyond the SM. What is unique to SUSY, beyond leading to a set of consistent and
completely formulated models, as, for example, the MSSM, is that this theory can potentially work up to the GUT energy scale.
In this respect it is the most ambitious model because it describes a computable framework that could be valid all the way
up to the vicinity of the Planck mass. The SUSY models are perfectly compatible with GUT's and are actually quantitatively
supported by coupling unification and also by what we have recently learned on neutrino masses. All other main ideas for going
beyond the SM do not share this synthesis with GUT's. The SUSY way is testable, for example at the LHC, and the issue
of its validity will be decided by experiment. It is true that we could have expected the first signals of SUSY already at
LEP, based on naturality arguments applied to the most minimal models (for example, those with gaugino universality at
asymptotic scales). The absence of signals has stimulated the development of new ideas like those of large extra dimensions
and "little Higgs" models. These ideas are very interesting and provide an important reference for the preparation of LHC
experiments. Models along these new ideas are not so completely formulated and studied as for SUSY and no well defined and
realistic baseline has sofar emerged. But it is well possible that they might represent at least a part of the truth and it
is very important to continue the exploration of new ways beyond the SM. New input from experiment is badly needed, so we all look forward to the start of the LHC.

I conclude by thanking the Organisers of this very inspiring Meeting: Guido Montagna, Oreste Nicrosini and Valerio Vercesi, for their kind invitation and great hospitality in Pavia.





%
%


\end{document}